\documentclass{appolb}
\usepackage{graphicx}

\usepackage[latin1]{inputenc}
\usepackage{graphics}
\usepackage{graphicx}
\usepackage{epsfig}
\usepackage{color}
\usepackage{longtable}
\usepackage{hyperref}
\usepackage{latexsym}
\usepackage{amsmath}
\usepackage{amsthm}
\usepackage{amsfonts}
\usepackage{amssymb}
\usepackage{dsfont}
\usepackage{bm}
\usepackage{graphics,psfrag}
\usepackage{graphicx,psfrag}
\usepackage[nice]{nicefrac}

\graphicspath{}
\newcommand{\be}{\begin{equation}}
\newcommand{\ee}{\end{equation}}
\newcommand{\bea}{\begin{eqnarray}}
\newcommand{\eea}{\end{eqnarray}}

\newcommand{\E}{\mathrm{e}}

\renewcommand{\imath}{\mathrm{i}}

\newlength{\mylenC}
\newlength{\mylenL}
\newlength{\mylenR}
\newlength{\mylenRm}
\begin{document}
%%%%%%%%%%%%%%%%%%%%%%%%%%%%%%%%%%%%%%%%%%%%%%%%%%
%%%%%%%%%%%%%%%%%%%%%%%%%%%%%%%%%%%%%%%%%%%%%%%%%%

\title{QCD with Two Light Dynamical \\Chirally Improved Quarks\thanks{Presented at ``Excited QCD 2013'', Bjelasnica Mountain, Sarajevo}
}

\author{Georg P.~Engel$^{1,2}$, C. B. Lang$^1$, \\Daniel Mohler$^{3}$, Andreas Sch\"afer$^4$
\address{
$^1$Institut f\"ur Physik, FB Theoretische Physik, \\ Universit\"at Graz, A--8010 Graz, Austria\\
$^2$Dipartimento di Fisica, Universit\`a di Milano-Bicocca and INFN, \\ sezione di Milano-Bicocca, I--20126 Milano, Italy \\
$^3$Fermi National Accelerator Laboratory, Batavia, Illinois, USA \\
$^4$Institut f\"ur Theoretische Physik, \\ Universit\"at Regensburg, D--93040 Regensburg, Germany}
}

\maketitle
\begin{abstract}
Results for the excited meson and baryon spectrum with two flavors of Chirally Improved sea quarks are presented. We simulate several ensembles with pion masses ranging from 250 to 600 MeV and extrapolate to the physical pion mass.  Strange quarks are treated within the partially quenched approximation. Using the variational method, we investigate the content of the states. Among others, we discuss the flavor singlet/octet content of Lambda states. In general, our results compare well with experiment, in particular we get very good agreement with the $\Lambda(1405)$ and confirm its flavor singlet nature. \end{abstract}
  
%%%%%%%%%%%%%%%%%%%%%%%%%%%%%%%%%%%%%%%%%%%%%%%%%%
\section{Introduction}
%%%%%%%%%%%%%%%%%%%%%%%%%%%%%%%%%%%%%%%%%%%%%%%%%%

Given the wealth of experimental data on hadron resonances \cite{Beringer:1900zz}, 
an ab-initio calculation starting from QCD would be highly desirable.
Although there is noticeable progress in calculating resonances on the lattice (for a review, see, e.g., \cite{Mohler:2012nh}), 
so far most lattice calculations are restricted to the calculation of the discrete spectrum in a finite box.
Towards larger volumes the discrete spectrum become denser, and the volume dependence is related to the phase shift of the resonance in the elastic region \cite{Luscher:1990ux,Luscher:1991cf}.
However, in finite volume and also for unphysically heavy pion masses the decay channels are often not
open or the corresponding phase space is small. 
This leads to energy levels in the finite system which are close to the resonance peak, at least for narrow resonances. 
One thus identifies such discrete energy levels with the masses of corresponding resonances. 
The present article is a further contribution in this spirit.
The results presented here have been published before in \cite{Engel:2011aa,Engel:2012qp,Engel:2013ig}, 
for recent reviews on hadron spectroscopy on the lattice, see, e.g., \cite{Bulava:2011np,Lin:2011ti,Fodor:2012gf}. 

%%%%%%%%%%%%%%%%%%%%%%%%%%%%%%%%%%%%%%%%%%%%%%%%%%
\section{Setup}
%%%%%%%%%%%%%%%%%%%%%%%%%%%%%%%%%%%%%%%%%%%%%%%%%%

We use the Chirally Improved (CI) fermion action \cite{Gattringer:2000js,Gattringer:2000qu}, 
which is an approximate solution to the Ginsparg-Wilson equation. 
A parametrization of a general fermion action is taken as ansatz, 
and the Ginsparg-Wilson equation is solved algebraically after truncation.
For the gauge sector we use the tadpole-improved L\"uscher-Weisz  
action \cite{Luscher:1984xn}. 
The lattice spacing $a$ is set at the physical pion mass for each value of the gauge coupling 
using a Sommer parameter $r_0=0.48$ fm, as discussed in \cite{Engel:2011aa}.
The energy levels are determined using the variational method
\cite{Luscher:1990ck,Michael:1985ne}. 
One constructs several interpolators $O_i$ for each set of quantum numbers 
and computes the cross-correlation matrix $C_{ik}(t)=\langle 
O_i(t) O_k(0)^\dagger\rangle$. 
The generalized eigenvalue problem
\begin{equation}
C(t) \vec u_n(t)=\lambda_n(t) C(t_0) \vec u_n(t)
\end{equation}
yields approximations to the energy eigenstates $|n\rangle$.
The exponential decay of the eigenvalues 
\begin{equation}\label{eq:eigenvalues}
\lambda_n(t)=\E^{-E_n\,(t-t_0)} (1+\mathcal{O}(\E^{-\Delta E_n(t-t_0)} ))
\end{equation}
is determined by the energy levels, where $\Delta E_n$ is the distance to
other spectral values. 
The eigenvectors tell about the content of the states 
in terms of the lattice interpolators. 
We simulate two CI light sea quarks and consider a valence strange quark. 
The strange quark mass is fixed by setting the $\Omega$-mass to its physical value. 
We generated seven ensembles of 200--300 gauge
configurations with pion masses in the range from 255 MeV to 596 MeV, lattice size $16^3\times 32$ and lattice 
spacing between 0.1324 and 0.1398 fm. For two ensembles with light pion masses
also lattices of size $12^3\times 24$ and $24^3\times 48$ were used to extrapolate to infinite volume. 
We use a large sets of interpolators in each quantum channel, combining
quark sources of different smearing width, different Dirac structure and 
different flavor structure. 
For the mesons, we also use derivative quark sources to enlarge the basis and to access spin 2 and exotic channels.
The interpolators are given explicitly in \cite{Engel:2011aa,Engel:2013ig}.
Albeit a large basis, it includes only quark--antiquark and 3-quark interpolators.
In principle, the sea quarks should provide overlap of these interpolators with 
meson--meson and meson--baryon states.
In practice, however, we cannot clearly identify such states.
A possible explanation may be a weak overlap with the used interpolators. 
The states may also be effectively included in the eigenstates, leading to a linear combination of exponentials in the eigenvalues. 
The present work is a continuation of \cite{Engel:2010my}, considering more ensembles, larger statistics and more observables. 

%%%%%%%%%%%%%%%%%%%%%%%%%%%%%%%%%%%%%%%%%%%%%%%%%%
\section{Results}
%%%%%%%%%%%%%%%%%%%%%%%%%%%%%%%%%%%%%%%%%%%%%%%%%%

\begin{figure}
\centerline{
\noindent\includegraphics[width=\mylenL,clip]{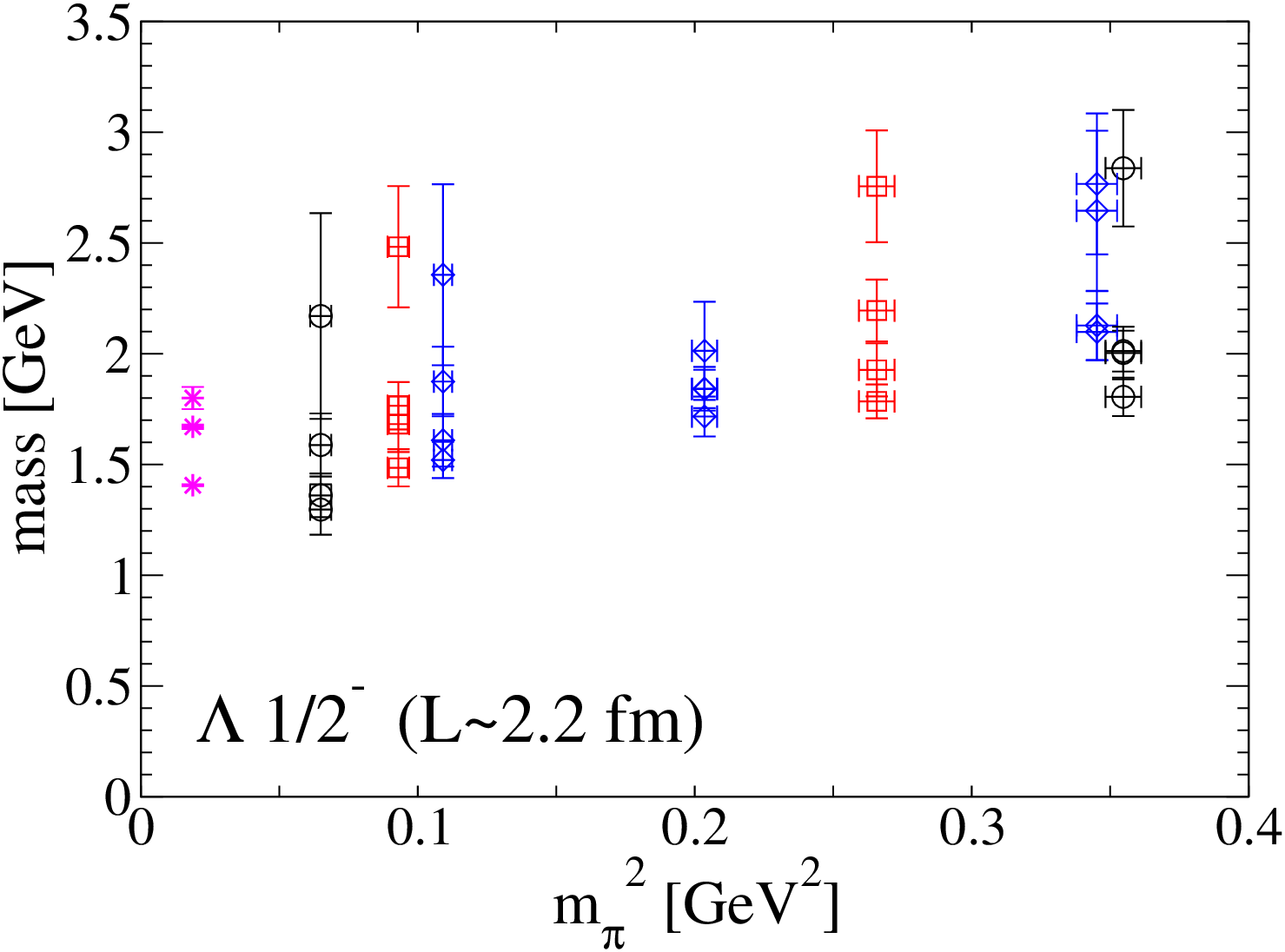}
\noindent\includegraphics[width=\mylenRm,clip]{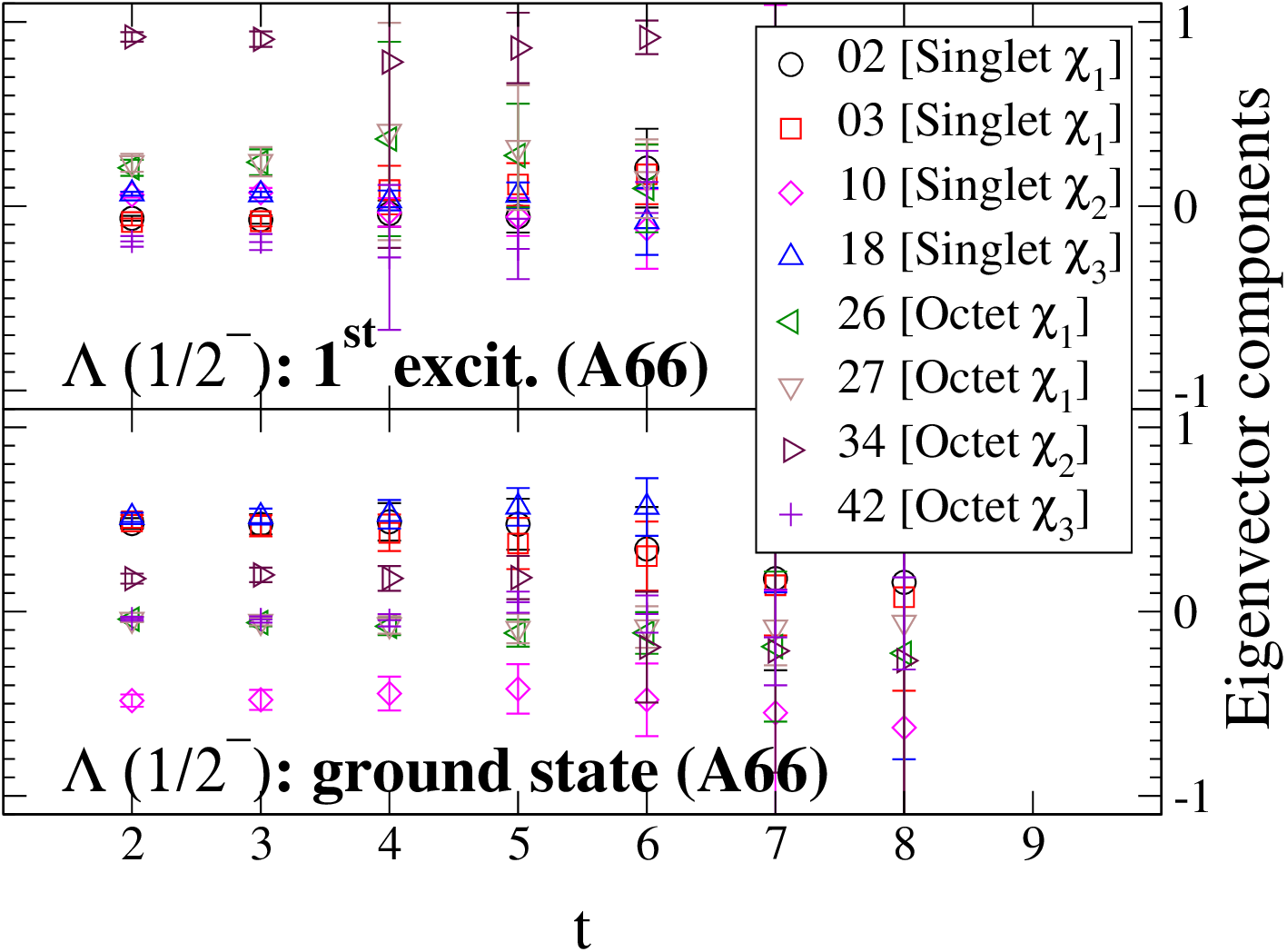}
}
\caption{Energy levels (left) and eigenvectors for the ground state and the first excitation for ensemble A66, $m_\pi\approx 255$ MeV, (right) for the baryon channel $\Lambda$ ($J^P=1/2^-$) in a finite box of linear size $L\approx 2.2$ fm.
On the left hand side, the magenta stars denote the experimental values \cite{Beringer:1900zz}, all other symbols represent results from the simulation. 
The ground state is dominated by singlet, the first excitation by octet interpolators. 
A considerable mixing of singlet and octet is observed (15-20\% for the ground state). 
}
\label{fig:lambda}
\end{figure}

\begin{figure}
\centerline{
\noindent\includegraphics[width=\mylenL,clip]{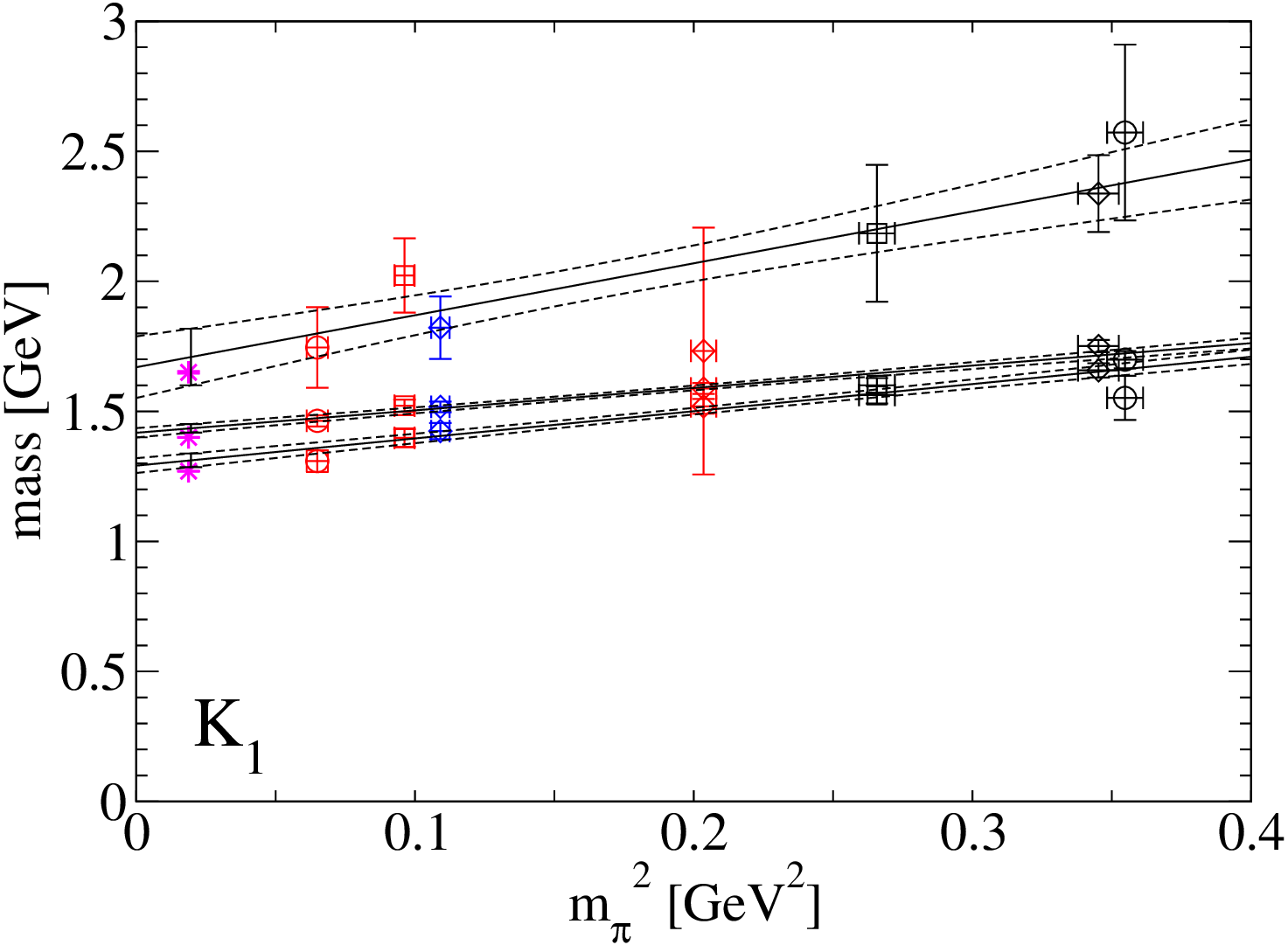}
\noindent\includegraphics[width=\mylenRm,clip]{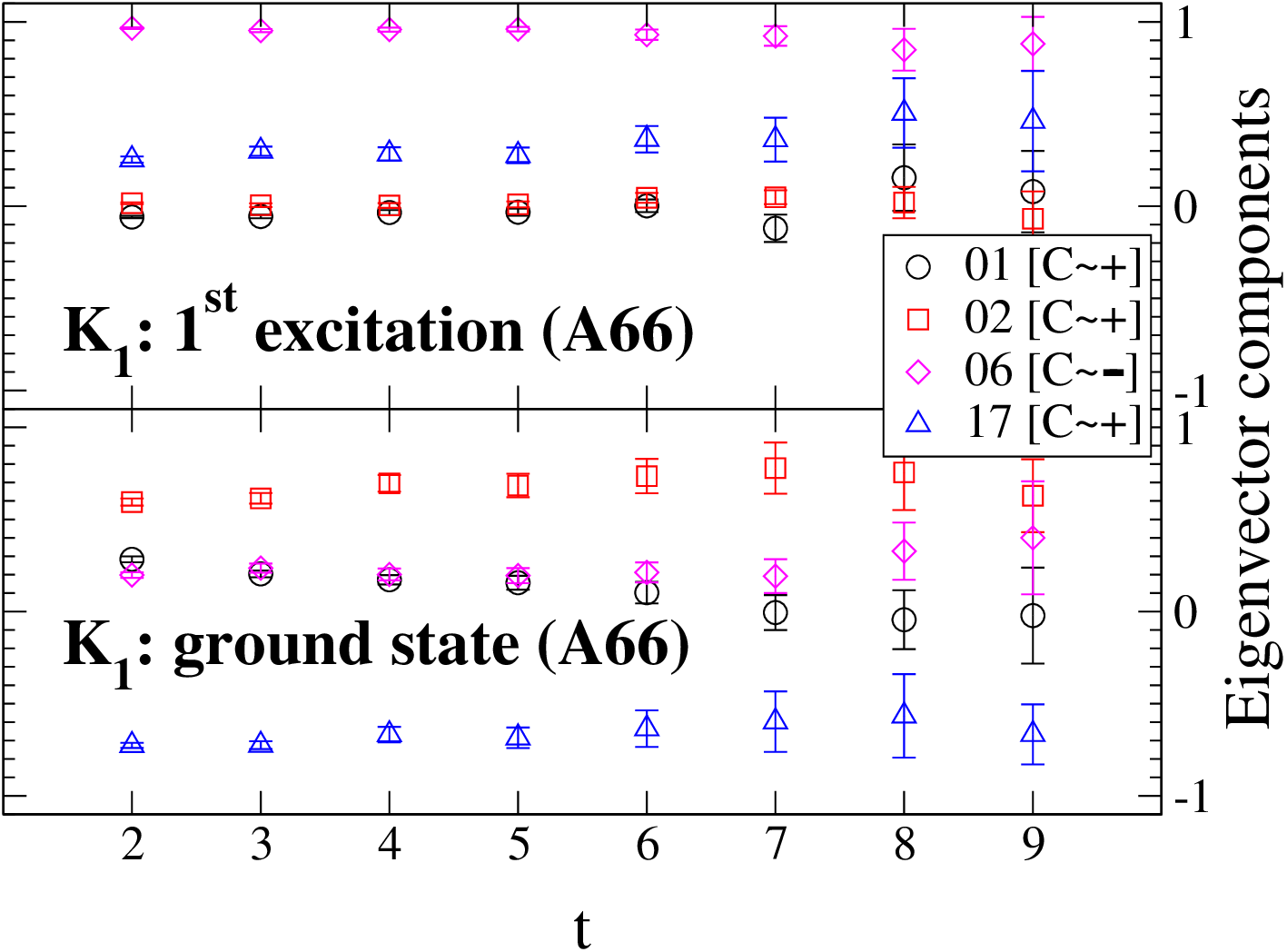}
}
\caption{Like Fig.~\ref{fig:lambda}, but for the strange meson channel $K_1$ ($J^P=1^{+}$) and including a fit of the energy levels linear in $m_\pi^2$.
The ground state (first excitation) is dominated by positive (negative) $C$-parity, which becomes a good quantum number in the limit of $SU(3)$ flavor symmetry. Note that mixing is observed in both states.}
\label{fig:strangemeson:1+}
\end{figure}

\begin{figure}[htb]
\centerline{
\noindent\includegraphics[width=\mylenL,clip]{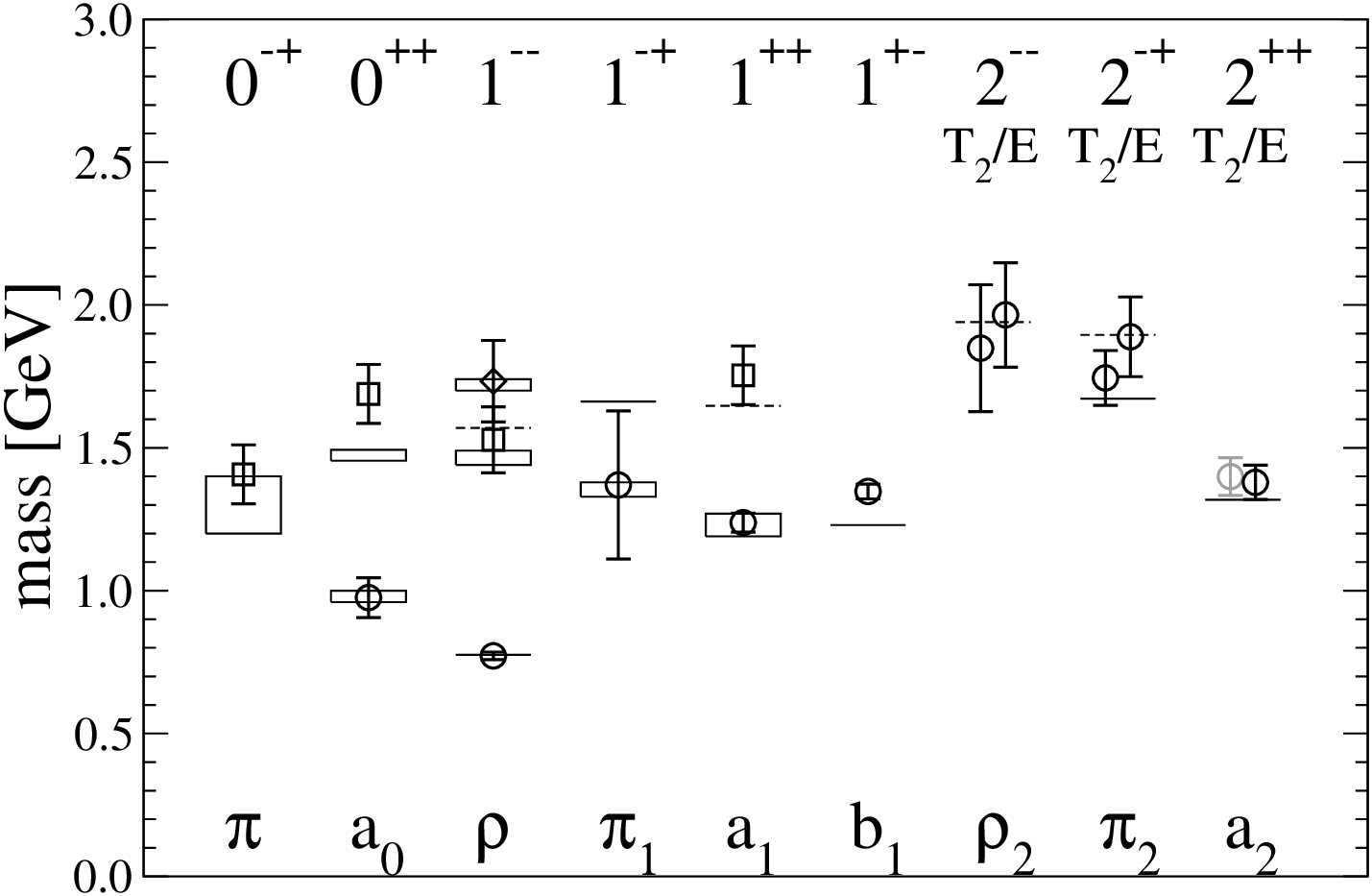}
\noindent\includegraphics[width=\mylenR,clip]{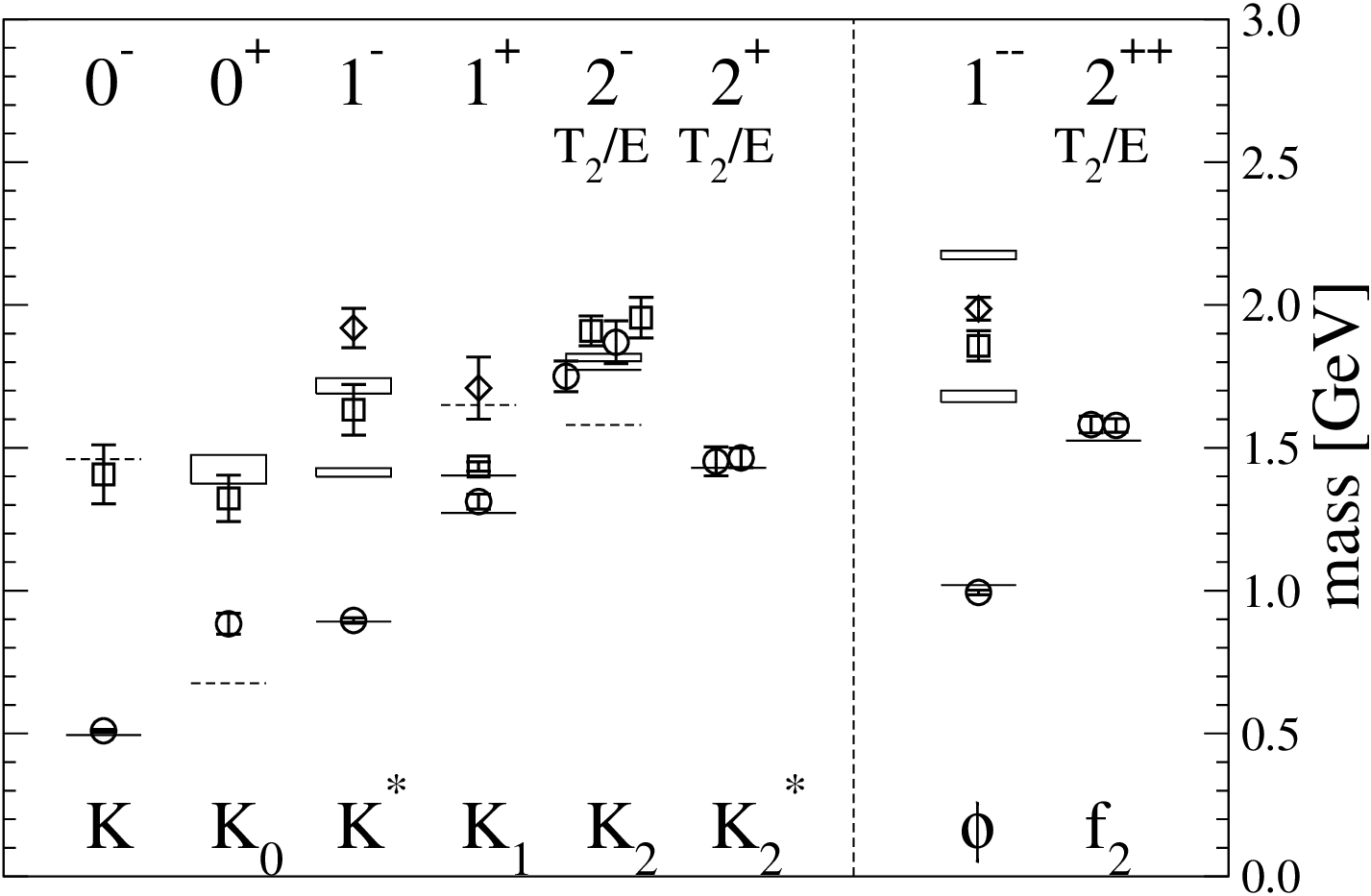}
}
\caption{
Energy levels for isovector light (left), strange and isoscalar (right) mesons in finite volume.
All values are obtained by chiral extrapolation linear in the pion mass squared.
Horizontal lines and boxes represent experimentally known states, 
dashed lines indicate poor evidence, according to \cite{Beringer:1900zz}.
The statistical uncertainty of our results is indicated by bands of $1\sigma$, 
that of the experimental values by boxes of $1\sigma$.
For spin 2 mesons, results for $T_2$ and $E$ are shown side by side.
Grey symbols denote a poor $\chi^2$/d.o.f.~of the chiral fits. 
Disconnected diagrams  are neglected in the isoscalars channels.
}
\label{fig:mesons_summary}
\end{figure}

\begin{figure}[htb]
\centerline{
\noindent\includegraphics[width=\mylenL,clip]{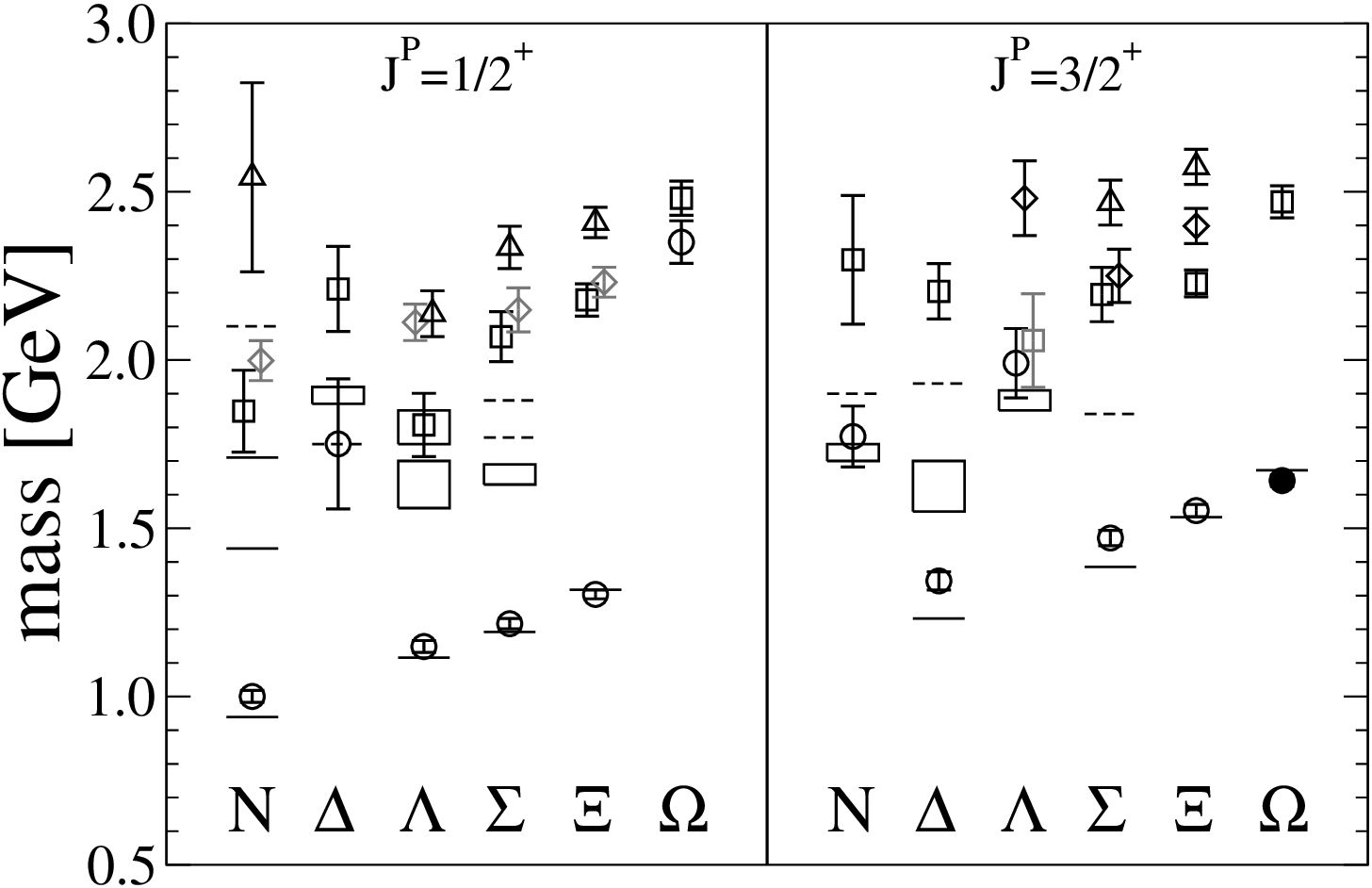}
\noindent\includegraphics[width=\mylenR,clip]{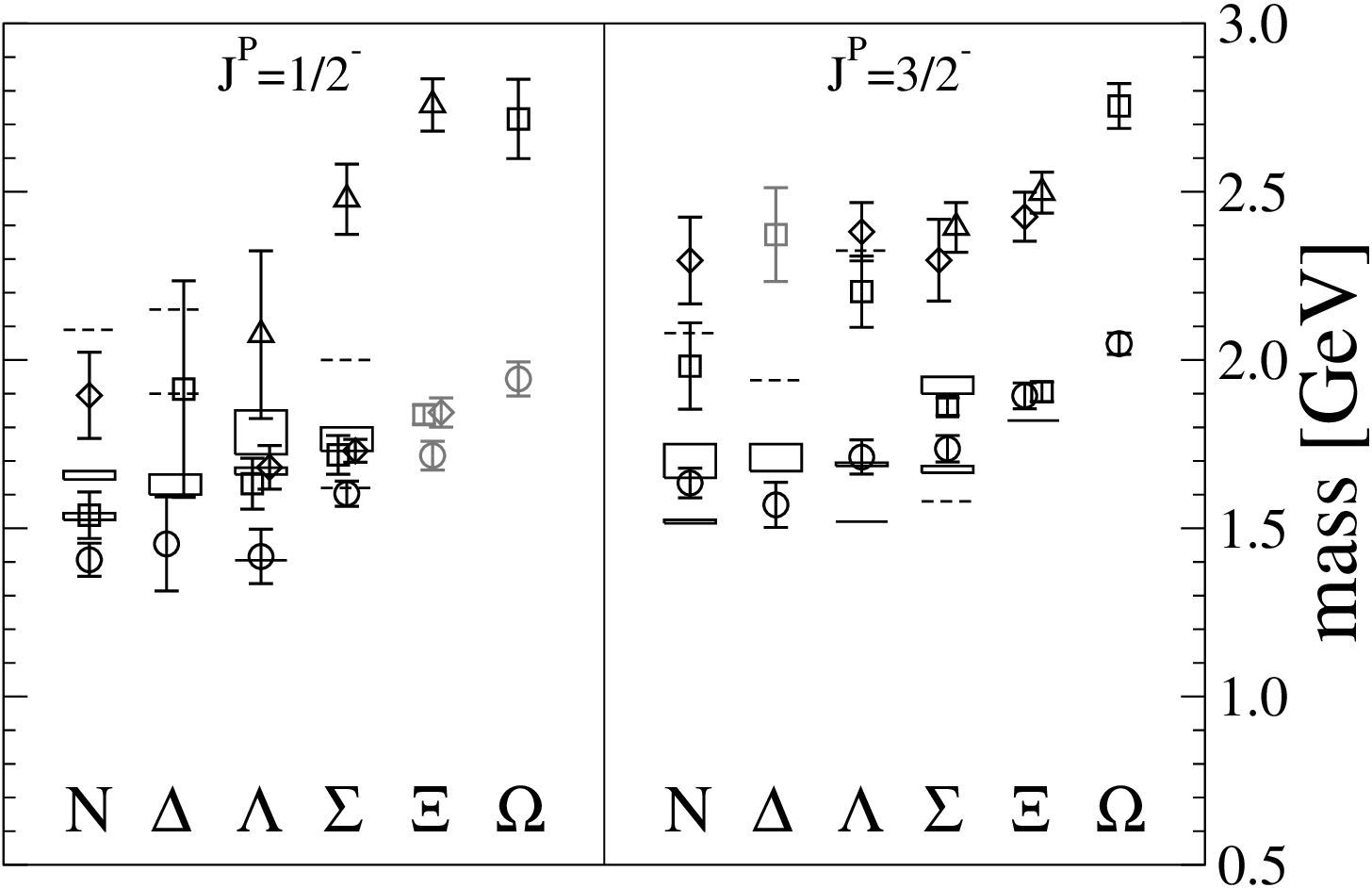}
}
\caption{
Like Fig.~\ref{fig:mesons_summary}, but for 
positive (left) and negative parity (right) baryons.}
\label{fig:baryons_summary}
\end{figure}

\begin{figure}[htb]
\centerline{
\includegraphics[width=\mylenC,clip]{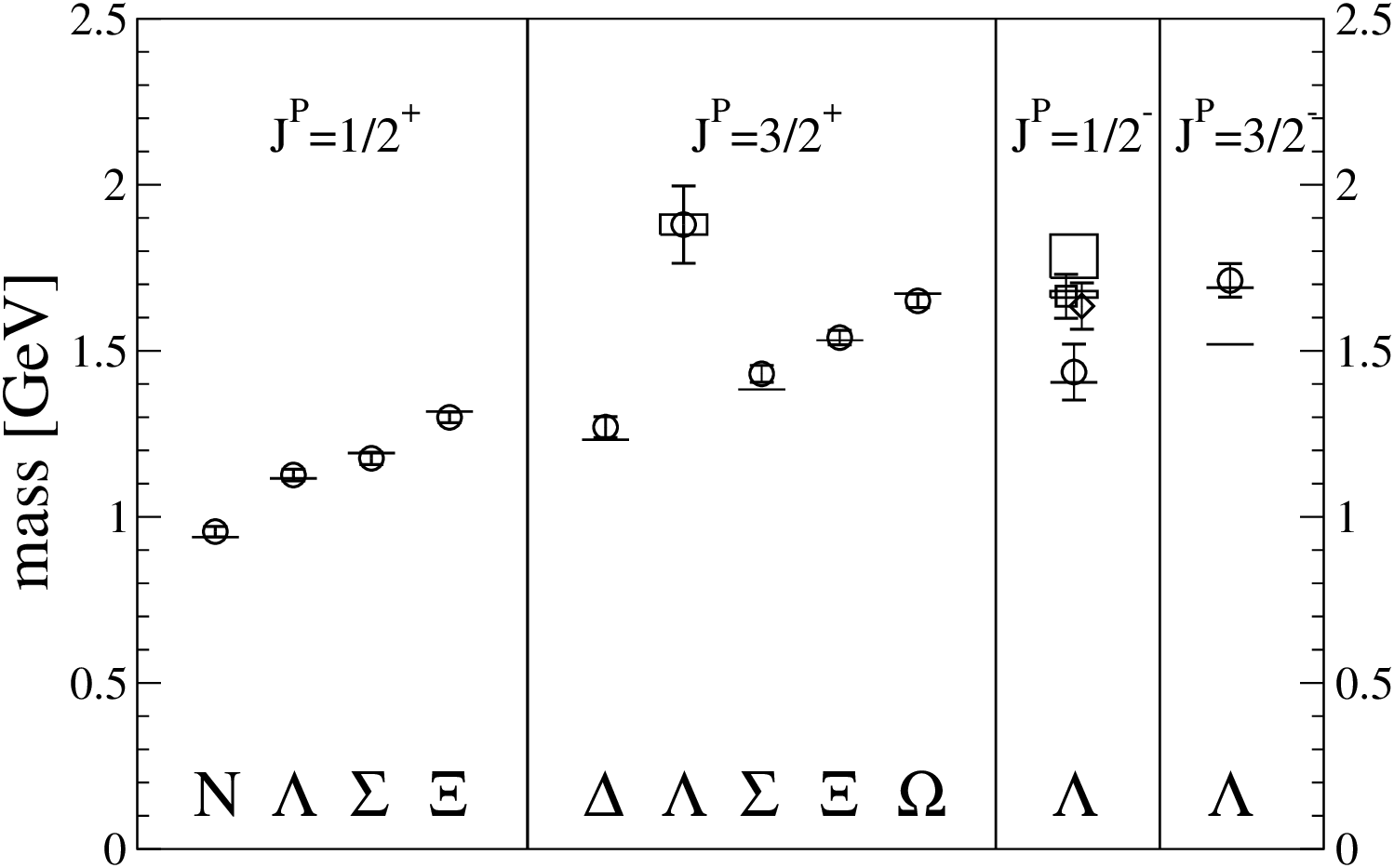}
}
\caption{
Like Fig.~\ref{fig:mesons_summary}, but for 
hadrons in the infinite volume limit.}
\label{fig:infvol_summary}
\end{figure}

We study the light and strange baryon channels $J=1/2,3/2$ in both parities and the light and strange meson channels $J=0,1,2$ in both parities and both $C$-parities.
Figs.~\ref{fig:mesons_summary} and \ref{fig:baryons_summary} show our results for the energy levels in finite volume after extrapolation to physical pion mass (linear in $m_\pi^2$).
In general, the low energy levels agree nicely with the experimental values, where available. 
As an example, we reproduce the prominent $\Lambda(1405)$. 
In Fig.~\ref{fig:lambda} (left hand side) we shown the corresponding
results fro the individual ensembles.
In some cases our results suggest  the existence of yet experimentally unobserved resonance states. 
However, some obtained energy levels mismatch experiment. 
One example is the first excitation in the nucleon ($J^P=1/2^+$) channel, which is considerably
higher than the Roper resonance. A possible interpretation is
a weak overlap of our interpolators with the physical state.
This may be also the case in the $\Lambda$ ($J^P=1/2^{+}$) channel, where the first excitation is dominated by singlet interpolators 
matching the $\Lambda(1810)$ (singlet in the quark model), while the Roper-like  $\Lambda(1600)$ (octet
in the quark model) seems to be missing. 
We study the systematic effects due to the choice of interpolators and fit ranges and 
we also perform
infinite volume extrapolations for several energy levels where the signal is good enough. 
Fig.~\ref{fig:infvol_summary} shows our results for the energy levels after extrapolation to infinite volume. 
The origin of remaining small deviations from experiment cannot be identified uniquely in our study. 
In general, however, our results in the infinite volume limit compare very well with experiment.
For the baryons, we analyze the flavor content by identifying the singlet/octet/decuplet contributions. 
For example, we find a dominance of singlet interpolators (mixing of 15-20\% with octet) for $\Lambda(1405)$, 
and a dominance of octet interpolators of the first excitation.
The corresponding eigenvectors are shown in Fig.~\ref{fig:lambda}.
For the strange mesons, we analyze the content of positive and negative $C$-parity, which would a good quantum number in the limit of $SU(3)$ flavor symmetry. 
As an example, we find a dominance of positive (negative) $C$-parity interpolators for the ground state (first excitation) in the strange meson channel $K_1$ ($J^P=1^{+}$) (see Fig.~\ref{fig:strangemeson:1+}), where also the lowest three energy levels agree well with experiment. 

%%%%%%%%%%%%%%%%%%%%%%%%%%%%%%%%%%%%%%%%%%%%%%%%%%
\section{Acknowledgments}
%%%%%%%%%%%%%%%%%%%%%%%%%%%%%%%%%%%%%%%%%%%%%%%%%%

We thank E.~Gamiz, Ch.~Gattringer, L.~Y.~Glozman, M.~Limmer, 
W. Plessas, H.~Sanchis-Alepuz, M.~Schr\"ock and V.~Verduci
for valuable discussions.  The calculations have been performed at the Leibniz-Rechenzentrum Munich and on clusters at the
University of Graz. We thank these institutions.
G.P.E.~and A.S.~acknowledge support by
the DFG project SFB/TR-55. 
G.P.E.~was partially supported by the MIUR--PRIN contract 20093BM-NPR. 
Fermilab is operated by Fermi Research Alliance, LLC under Contract No. De-AC02-07CH11359 with the United States Department of Energy.

%%%%%%%%%%%%%%%%%%%%%%%%%%%%%%%%%%%%%%%%%%%%%%%%%%
\providecommand{\href}[2]{#2}\begingroup\raggedright\endgroup
%%%%%%%%%%%%%%%%%%%%%%%%%%%%%%%%%%%%%%%%%%%%%%%%%%

%%%%%%%%%%%%%%%%%%%%%%%%%%%%%%%%%%%%%%%%%%%%%%%%%%
%%%%%%%%%%%%%%%%%%%%%%%%%%%%%%%%%%%%%%%%%%%%%%%%%%
\end{document}